\documentclass{elsart}
\usepackage{amssymb,amsfonts}
\usepackage{epsfig}
\begin{document}
\begin{frontmatter}
\title{Transparency of 0.2$\%$ GdCl$_3$ Doped Water in a Stainless Steel Test Environment}

\author[lsu]{W. Coleman\thanksref{corresponding}},
\author[llnl]{A. Bernstein},
\author[llnl]{S. Dazeley},
\author[llnl,ucd]{R. Svoboda}

\address[lsu]{Department of Physics and Astronomy, Louisiana State 
University, Baton Rouge, LA 70803, USA}
\address[llnl]{Advanced Detectors Group, AP-Div, Lawrence Livermore National Laboratory, 
Livermore, CA 94550, USA}
\address[ucd]{Department of Physics, University of California Davis, 
Davis, CA 95616, USA}
\thanks[corresponding]{Corresponding author: Department of Physics and Astronomy, Louisiana State 
University, Baton Rouge, LA 70803, USA ,
Tel: +1 225 578 2261, Fax: +1 225 578 5855, e-mail wcolem2@lsu.edu} 

\begin{abstract}The possibility of neutron and neutrino detection using water Cerenkov detectors doped with gadolinium holds the promise of constructing very large high-efficiency detectors with wide-ranging application in basic science and national security.  This study addresses a major concern regarding the feasibility of such detectors: the transparency of the doped water to the ultraviolet Cerenkov light.  We report on experiments conducted using a 19-meter water transparency measuring instrument and associated materials test tank.  Sensitive measurements of the transparency of water doped with 0.2$\%$ GdCl$_{3}$ at 337nm, 400nm and 420nm were made using this instrument.  These measurements indicate that the use of GdCl$_{3}$ in stainless steel constructed water Cerenkov detectors is problematic.
\end{abstract}

\begin{keyword}
Water Cerenkov Detercors; Neutron; Neutrino, DSNB, Transparency; Gadolinium\\
PACS: 29.40.Ka 
\end{keyword}
\end{frontmatter}

\section{Introduction}

Water Cerenkov Detectors (WCDs) have played a significant role in 
advancing the field of particle physics.  The combination of large mass coupled with low cost
led to their use in the detection of cosmic ray\cite{cr} , solar\cite{solar1,solar2}, and accelerator neutrinos\cite{accel} and resulted in some of the most definitive evidence for neutrino
oscillations.  In addition, WCDs have been used to set limits on proton decay\cite{prot1,prot2}
and to detect neutrinos from supernovae\cite{sn}.

The principle of operation of current detectors is rather simple.  Water has a refractive index (n) of roughly 1.33 and so charged particles moving at a speed greater than  c/n (where $c$ is the speed
of light in a vacuum) will emit Cerenkov photons at an angle $\theta 
= \arccos$ (1/n) to their direction of travel. These photons have a characteristic $1 / \lambda^{2}$ 
spectrum, which means that much of the detectable light is emitted in the ultraviolet. Pure water happens to have a `transparency window' with
attenuation lengths greater than 50 m in the spectral region from 
about 320 to 480 nm.  This is well-matched to the Cerenkov spectrum and the sensitivity of 
bi-alkali photomultiplier tubes.  Figure 1 shows the attenuation coefficient ($\alpha$) plotted against wavelength for purified water contained in the 50K ton Super-Kamiokande neutrino detector\cite{SK}.

Unfortunately, water from typical public potable water supplies has a 
much shorter attenuation length in the UV, often 5-10 meters or less.  The reason for this is not 
well-understood.  Nevertheless, standard water purification systems that remove dissolved solids have empirically been found to solve this problem.  Thus all operating
water Cerenkov neutrino detectors have an associated water treatment plant.  This component is a significant percentage of the construction expense.  In addition, it is found that the water in the detector will deteriorate in transparency - even under a nitrogen atmosphere and in a stainless steel or high-density polyethylene (HDPE) tank.  Thus the water treatment plant must be run continuously to maintain detector water quality and stability.  This is often the largest operational expense of WCDs.

\begin{figure}
\begin{center}
\epsfig{file=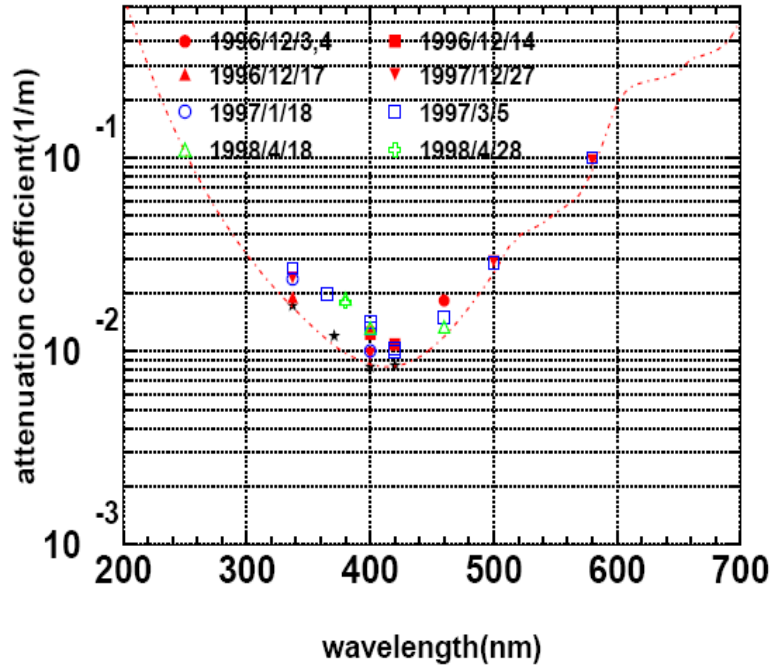,height=10cm,width=0.8\textwidth}
\end{center}
\caption{Water attenuation coefficient ($\alpha$) plotted as a function of wavelength together with the prediction model used in the Super-Kamiokande Monte Carlo simulation (dashed-dotted lines)~\cite{SK}}
\label{f:transparency}
\end{figure}

Recently Super-Kamiokande has published limits on a search for `relic supernovae'\cite{rsn}.  While nearby supernovae are rare, the expected rate integrated over the entire universe is
substantial and is predicted to produce a diffuse supernova neutrino background
[DSNB].  This flux, if observable, could provide information about
stellar collapse and nucleo-synthesis, and the rate of stellar formation as a function of red
shift.  These limits are still a factor of 4-5 away from the most optimistic 
models of stellar formation.  In this case, most of the expected interactions come
from $\overline{\nu}_{e}$ interactions with hydrogen in the water due to the
so-called `inverse beta decay' reaction:

\begin{center}
$p + \overline{\nu}_{e} \rightarrow n + e^{+}$
\end{center}

There are significant backgrounds for this reaction from atmospheric neutrino
interactions in the same range.  Many of these are from muon neutrino interactions
that produce a muon below the Cerenkov threshold which subsequently decays 
to produce a Michel electron or positron in the middle of the SN neutrino energy range.  In
reactor neutrino experiments using this same interaction, the capture of
the daughter neutron and subsequent gamma emission are used as an effective tag to 
discriminate against background. Unfortunately, the 2.2 MeV gamma from capture on hydrogen in the water results in relatively little Cerenkov emission.  Thus there is interest in dissolving in the water nuclei with large capture cross section and higher energy gamma emission.

Use of CdSO$_4$ in water has been previously discussed \cite{CdSO4}.  Beacom and Vagins\cite{BeaVag} have also suggested the use of GdCl$_3$ in order to boost the capture gamma energy via an 8 MeV gamma cascade.  They pointed out that the addition of only small amounts of GdCl$_3$ (0.2$\%$ by weight) would result in about 90$\%$ of the neutrons being captured on Gd.  This would reject a substantial part of the atmospheric neutrino background and greatly improve the sensitivity of Super-Kamiokande for DSNB events.  This would also be advantageous to separate `cooling' from `neutronization' neutrinos should we be fortunate enough to detect a supernova in our own galaxy.  Thus there is substantial motivation to use chemical additives in both Super-Kamiokande
and in possible future megaton scale detectors.

In addition to neutrino detection, WCDs are now being considered as active water
shields for some next-generation Dark Matter (DM) / Weakly Interacting Massive Particle (WIMP) detectors.  Such particles could comprise all or part of the DM in the universe.  Experimental searches for WIMPs have been going on for 20 years with ever increasing sensitivity.  Recently,
liquid nobel gas detectors have shown great promise for further extending the search
one or two more orders of magnitude.  In these detectors, WIMPs would show up
as low energy nuclear recoil events in a heavy liquid target such as xenon or argon.

Nuclear recoil events produced by the interaction of non-thermal neutrons in the target are a
significant background for these experiments.  A transfer via elastic scatter,
followed by neutron escape from the detector, can mimic a WIMP recoil.  While tagging
of a large fraction of muon-induced events is possible with a simple muon veto system, direct
rejection of un-vetoed or non-muon initiated nuclear recoils is more difficult.  Pulse Shape
Discrimination (PSD) is of no help here, since single neutron-induced recoils are
essentially identical in light and charge deposition and timing to WIMP-induced ones.  A dangerous source of background are radioactive U/Th contaminants in the materials of the
detector itself, with a major worry being ($\alpha$,n) reactions on boron and silicon in the PMTs. 

One way to reduce internal neutron backgrounds is to efficiently tag those neutrons that
leave the central detector sensitive volume.  Estimates from the LUX collaboration show that
this could potentially reduce such backgrounds by an order of magnitude.  Thus there is great interest in gadolinium doping of water for DM searches.

There are three potential problems associated with gadolinium doping of water:

\begin{enumerate}
\item The additive might reduce the transparency of water in the UV, seriously
reducing detector sensitivity.
\item The additive might induce corrosion that would affect the mechanical strength
of the detector components.
\item Corrosion product ions might absorb light in the UV, reducing detector sensitivity.
\end{enumerate}

Thus before any of these experiments can be done it is crucial to understand the effect
of the gadolinium dopant on the detector itself to avoid the `law of unintended
consequences.'

One of the least expensive and most readily available gadolinium compounds is
gadolinium chloride (GdCl$_{3}$).  It is highly soluble and known to be relatively
benign environmentally and from a health safety point of view.  In this paper, we report
the first results from a three year study on the effect of GdCl$_3$
on detector materials and water transparency relevant to existing and future
water Cerenkov detectors.

\section{Experiment Description}
\subsection{Experimental Apparatus}
Figure 2 is a schematic of the LLNL Water Cerenkov Transmission Facility (WCTF).  It consists of a water purification system (water resin demineralizers, conductivity sensor, 0.22$\mu$m and 5$\mu$m filters, polypropylene transfer piping, a mixing tank, and 1.27cm diameter 304L grade stainless steel transfer piping which is used to fill the Light Transmission Arm (LTA).  The LTA is a 9.6m, 20.3cm diameter 304L stainless steel pipe which can be filled with pure or doped water.  During the process of mixing the GdCl$_3$, nitrogen gas is bubbled through the mixing tank to reduce the dissolved oxygen content in the system.

\begin{figure}
\begin{center}
\epsfig{file=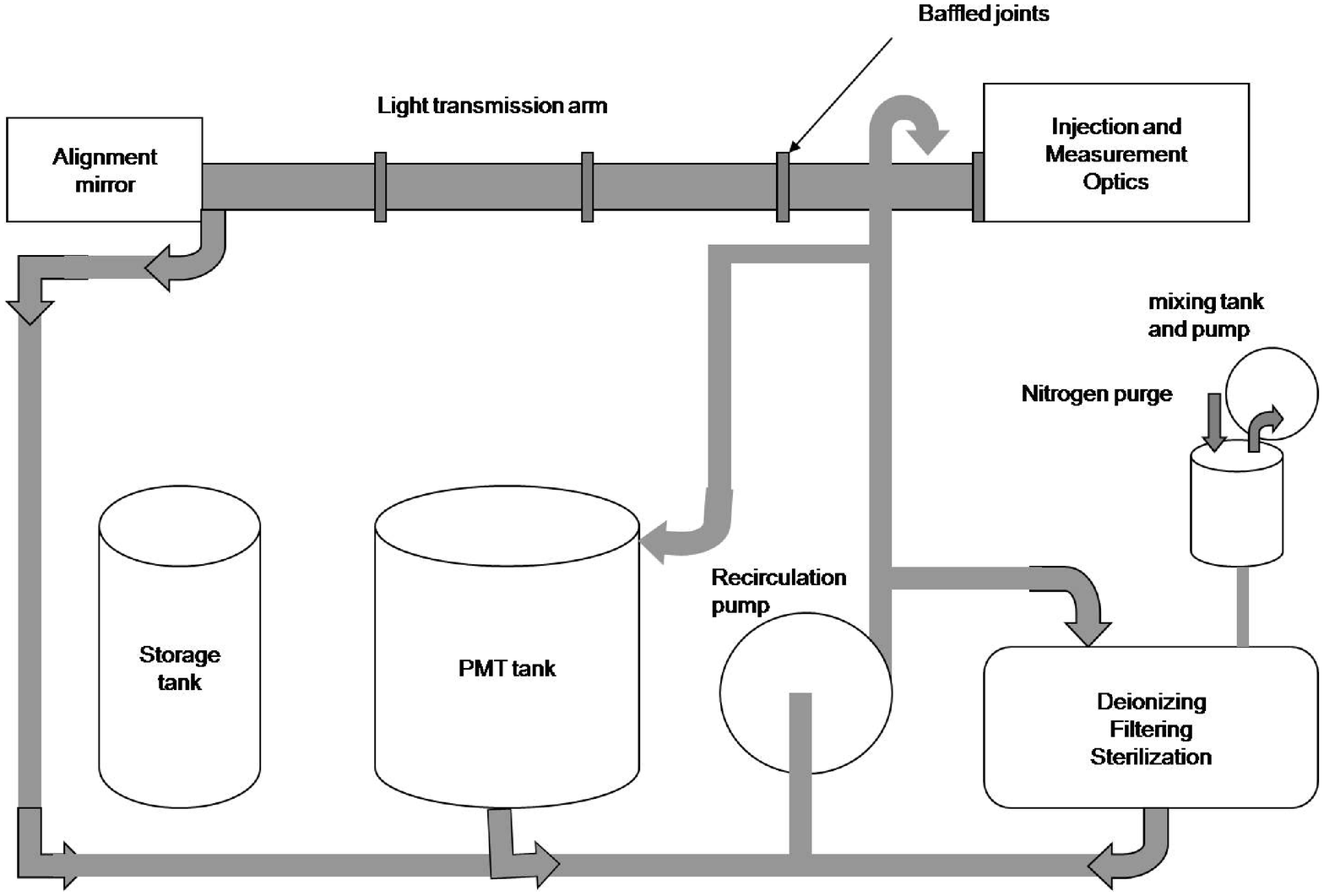,height=10cm,width=0.8\textwidth}
\end{center}
\caption{The LLNL Water Cerenkov Transmission Facility.  The system can produce ultra$-$pure  water and inject GdCl$_3$ via a mixing tank.  The water is also circulated through 5 and 0.2 micron filters and an ultraviolet sterilizer to suppress bacterial growth.  The water system services both the light transmission arm and the materials test tank, containing two 50-cm diameter PMT’s on loan from Super-Kamiokande.}
\label{f:picture30}
\end{figure}

The water purification system is capable of providing high purity water (Resistivity $>$ 17.5M$\Omega$ cm) to the stainless steel LTA at a flow rate of 5.7$-$11.4 lpm.  This resistivity value corresponds to total dissolved solids of $\lesssim$ 40 ppb.  The mixing tank, LTA and transfer piping contain a total of about 570 liters and the turn$-$over time (at a typical flow rate of 9.5 lpm) is approximately one hour.

The effects of exposure to GdCl$_3$ on materials used in Super-Kamiokande was previously reported in \cite{mater}.  Additional testing was conducted by placing two 50.8cm Super-Kamiokande PMTs in a 2470 liter stainless steel storage tank filled with GdCl$_3$.  The results of these tests will be reported at a later date.

The system is capable of testing the attenuation length of water at three wavelengths. The output of an L.S.I. VSL-337ND-S pulsed nitrogen laser can produce a 337nm beam directly.  In addition, the N$_{2}$ laser output can pump organic dyes.  The two dyes used were Stilbene-420 which produces a 420nm pulse and PPBO which produces a 400nm pulse. 

Figure 3 shows the detector optics arrangement used for the conduct of the transparency testing.  The optics are arranged on optical tables and enclosed in two light-tight boxes at both ends of the LTA.  The LTA is closed at both ends by 0.8cm thick ultra$-$violet light transmitting (UVT) acrylic windows.  Except for a 3cm $\times$ 20cm slit, the two acrylic windows at each end of the LTA are covered by black plastic to prevent unwanted external light from entering.  The LTA contains four identically sized baffles spaced at intervals in the interior to remove scattered light.  The L.S.I. laser generates a 337nm 4ns pulse at approximately 2 Hz.  The laser beam is collimated by two apertures of 6.3mm and 3.0mm diameter.  The beam is then split by a UV non-polarizing cube beam-splitter.

One part of the beam - designated the primary (P) beam - is sent directly into a coated integrating sphere and then through a UV transmitting liquid lightguide (Lumatec Series-250) to a white rectangular integrator which houses a mu-metal shielded PMT (Hamamatsu H3378-50).  The other beam - designated the reflected (R) beam - is transmitted directly through the UVT acrylic window into the LTA.  The R beam exits the far end of the LTA into a light-tight enclosure which houses a 5.08cm diameter UV mirror (Newport Optics BBDS-PM-2037-C $>$ 99$\%$ reflectivity).  This mirror is aligned to reflect the R beam back through the LTA and into the laser enclosure.  After re-entering the laser enclosure, the R beam is transmitted through a 10$\%$ transmitting neutral density filter (ND) into a spherical integrator (for 337nm) or an acrylic `light-box' containing a solution of LUDOX (HS-40 12nm diameter colloidal silica) and pure water (for 400nm and 420nm).  The R beam is then transmitted through a separate Lumatec lightguide into the rectangular enclosure housing the PMT.  Use of the same PMT to measure both the reflected and primary pulses allows us to cancel PMT gain variations.
\begin{figure}
\begin{center}
\epsfig{file=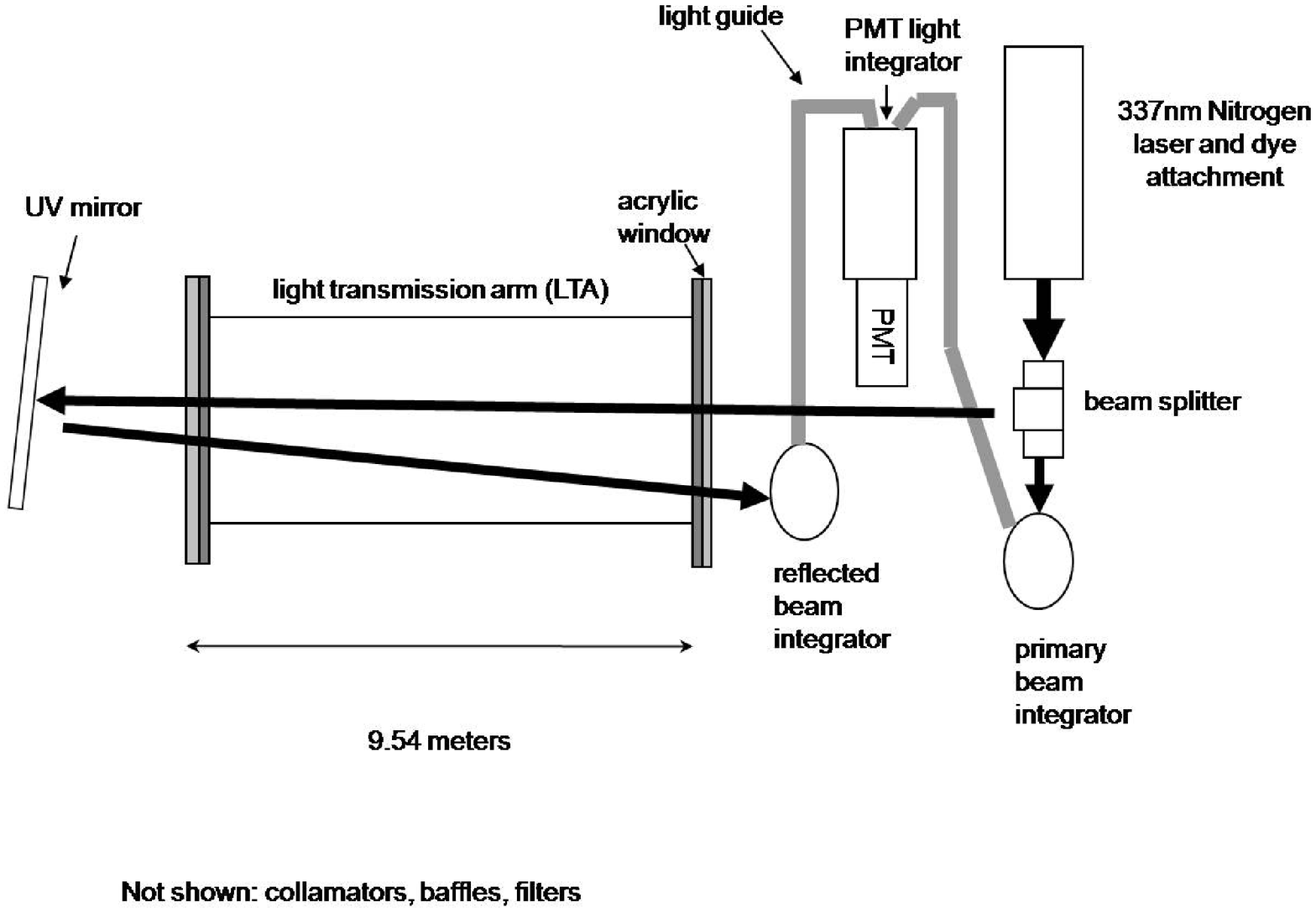,height=10cm,width=0.8\textwidth}
\end{center}
\caption{Schematic of injection and reflected beam optics.  A laser beam is split into two, the the primary beam is directed via an integrator to the PMT.  The reflected beam traverses the LTA and is reflected into a second integrator, where it is collected and sent to the PMT.  Delay time between the primary and reflected is $\approx$ 90ns, sufficient to cleanly separate them in time.}
\label{f:apparatusc}
\end{figure}

\subsection{Description of Measurements}

The change in the attenuation length of the GdCl$_3$-water solution can be determined by measuring the change in the ratio of the R laser pulse to the P laser pulse over time.

Let 
\begin{equation}
I = I_o \exp(- \alpha L) 
\end{equation}
where $I$ is the intensity of the pulse at the distance $L$ through the LTA, $I_o$ is the initial intensity and $\alpha$ is the attenuation coefficient in m$^{-1}$.  Assuming that P, R are proportional to I$_o$ and I respectively, then: 

\begin{equation}
P = aI_o  ;   R = bI
\end{equation}

where $I$ is the intensity of the reflected pulse and $a$ and $b$ are constants of proportionality.  We define the ratio:

\begin{equation}
\rho = \frac {R} {P} = \frac {b} {a} \exp(- \alpha L)
\end{equation}

Letting $\rho_1$ be the ratio in pure water and $\rho_2$ be the ratio after adding GdCl$_3$, then:
\begin{equation}
\Delta \alpha = {\alpha}_2 - {\alpha}_1 = \frac {1} {L} \ln \frac {\rho_1} {\rho_2}
\end{equation}
where the unknown constants $a$ and $b$ cancel.  The uncertainty in $\Delta \alpha $ is then given by:
\begin{equation}
\sigma_{\Delta \alpha} = \sqrt{ (\frac {\sigma_{\alpha1}}{\rho_1})^2 +  (\frac {\sigma_{\alpha2}}{\rho_2})^2 }
\end{equation}
Typical values for pure water $\alpha$ at UV wavelengths are 0.01 m$^{-1}$ $-$ 0.02 m$^{-1}$.  

\subsection{Methods}
The P and R beams were observed as two pulses well separated by about 90ns on a Tektronix DPO 4034 digital oscilloscope.  Figure 4 shows a typical trace for a double pulse at 337nm.  To determine the intensities, we integrate over the area of the single P and R pulses and then calculate the ratio $\rho =$ R / P.  To obtain $\rho$, we find the mean and variance of one-hundred separate pulses taken over a duration of about three minutes.  This set of 100 pulses will be referred to as a `measurement'.

In order to understand the response of the detector to changes in transparency caused by the addition of GdCl$_{3}$, we conducted a series of `control' experiments by adding only pure water to the LTA and performing measurements identical to those to be performed with the GdCl$_{3}$ water solution.  

Before each measurement, the reflected beam was aligned to the same point at the entrance of the R beam light integrator and re-checked for alignment after completing each measurement.  Typically it took about 10 minutes to take all readings between alignment checks.  If the beam moved more than about 2cm from the initial point of alignment, the measurement was discounted and a new measurement was taken after the apparatus had come to equilibrium.

To prove the stability of our detector, another control experiment conducted was to circulate pure water through the water purification system and the LTA, secure the recirculation and then take measurements of $\rho$ over a two week period while it remained sitting in the LTA (i.e. no circulation).  These measurements are shown as a function of time in Figure 5 and indicate a 12.5$\%$ decrease in $\rho$ over the period of more than two weeks.  While the cause of the loss of transparency is unclear, it is of interest to note that the $\sim$ 1.0$\%$ per day fall off in transparency is less than that typically observed in SK with the recirculation turned off, even though the stainless steel surface to volume ratio of our apparatus is almost 200 times larger.

\begin{figure}
\begin{center}
\epsfig{file=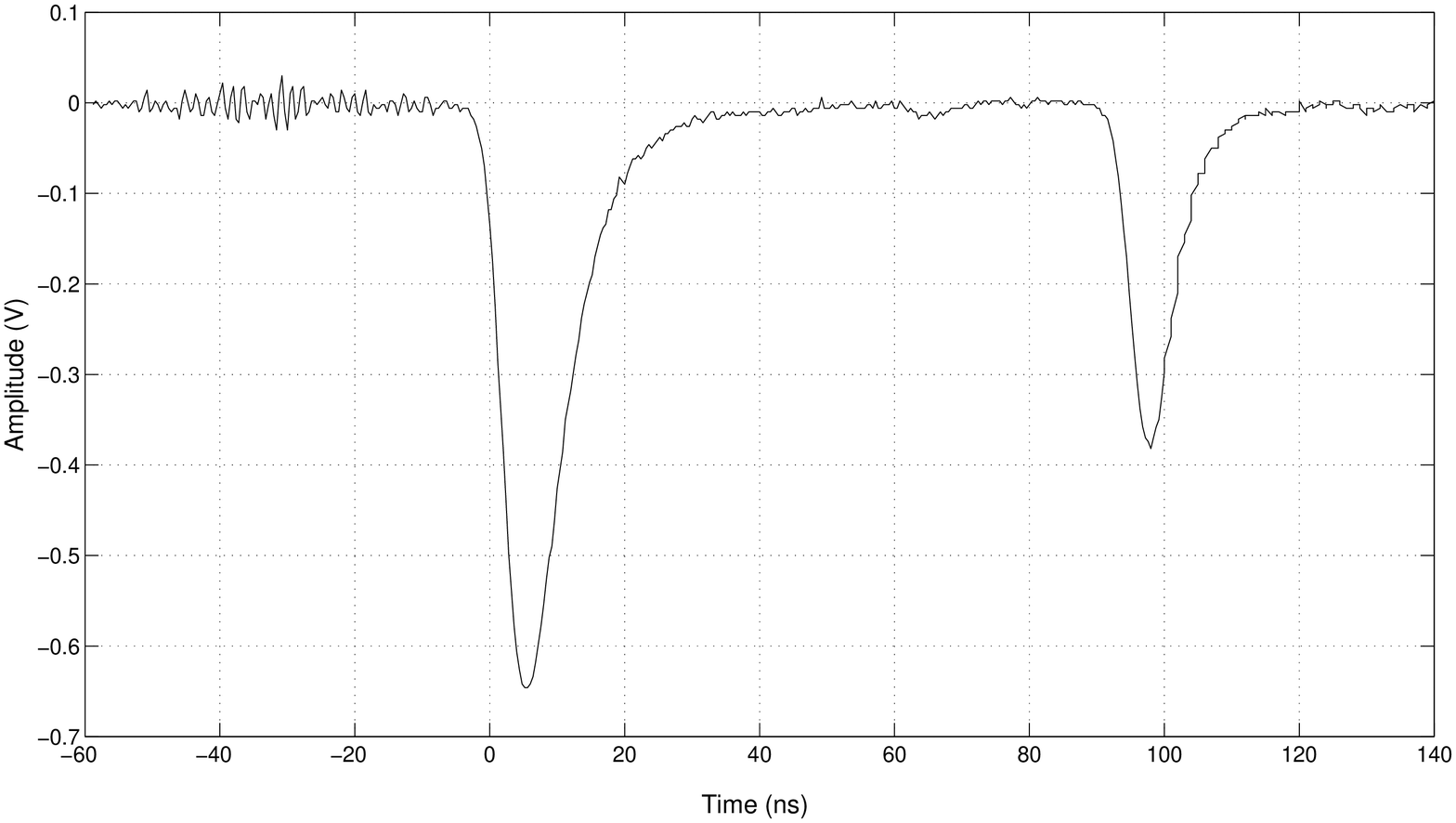,height=10cm,,width=0.8\textwidth}
\end{center}
\caption{Oscilloscope output for a typical P and R pulse at 337nm in pure water.  The left and right pulses are the single PMT responses to the P and R beams respectively.  The delay time between the primary and reflected is $\sim$ 90ns, which is sufficient to cleanly separate the pulses in time.  The time separation of the two pulses is due to the extra distance traveled by the R beam in water and air ($\Delta t = (19.25 $ m)($\frac {1.34} {3 \times 10^8 m/s})$  = 86ns + $\sim$ 3ns for air).  The small amount of ripple is due to R.F. pick-up from the laser fire.}
\label{f:337oscope}
\end{figure}

\begin{figure}
\begin{center}
\epsfig{file=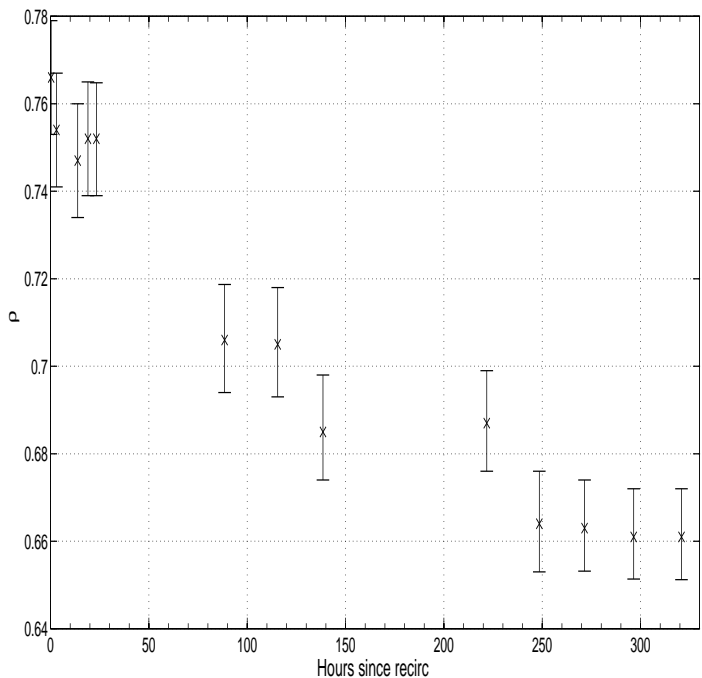,height=10cm,width=0.8\textwidth}
\end{center}
\caption{$\rho$ of pure water measured over approximately 14 days at 337nm.  Recirculation of the water through the system was turned off at t=0.  From this point, the water remained undisturbed in the LTA and $\rho$ decreased at the rate of $\sim$ 1$\%$ per day. }
\label{f:2inchstability_plot}
\end{figure}

For the 337nm and 420nm measurements, the amplitudes of both P and R were approximately equal for the pure water measurements.  This condition ensured that the response of the detector to both the P and R pulses is the same. However, this condition was not met for the case of the 400nm measurements due to the poor efficiency of the PBBO dye.  Because of this fact, an independent alignment of the optical system at 400nm was not possible.  As a consequence, the alignment uncertainty is greater for 400nm than the other two wavelengths as discussed in the section on uncertainties.

\section{Uncertainty}

Our measurements are affected by four significant sources of uncertainty.  The first of these is associated with the variation in the pulse-to-pulse measurement of $\rho$ over short time intervals due to thermal turbulence in water and the the vibrational motion of optical components.  To quantify this uncertainty, we take the average of the measured 100 pulse variation at each wavelength for each measurement of $\rho$ in pure water.  The values determined are:  for 337nm: $\pm 0.6\%$; for 400nm: $\pm4.4\%$; and for 420nm: $\pm2.0\%$.

A second uncertainty was associated with reproducibility of the the R beam alignment.  This uncertainty was quantified by conducting alignments of 10 randomly `misaligned' beams in rapid succession and obtaining the variations in the measurements of $\rho$.  Based on these measurements, the following estimates for the uncertainty were obtained: for 337nm: $\pm1.0\%$; and for both 400nm and 420nm: $\pm2.0\%$.

Another uncertainty is associated with the linearity of the system response to changes in light transparency.  In this report no measurement of $\rho$ below 45$\%$ were made so  only the detector response above this value is addressed.  To determine the detector linearity uncertainty, UV transmitting ND filters were inserted between the beam-splitter and the LTA and the measured $\rho$ compared to the expected light transmission.  The transmission of each of the ND filters was measured using a calibrated monochromometer to better than 0.5$\%$.  Figure 6 shows the normalized $\rho$ value plotted against the filter transmittance.  In this case, `normalized' means that the $\rho$ value with no filter present is taken to be 1.  The uncertainty in the light transmission linearity is conservatively taken to be the difference in the slope of a fitted line from 1. The corresponding uncertainties are: $\pm0.8\%$ for 337nm; $\pm1.1\%$ for 400nm and $\pm1.6\%$ for 420nm.

\begin{figure}
\begin{center}
\epsfig{file=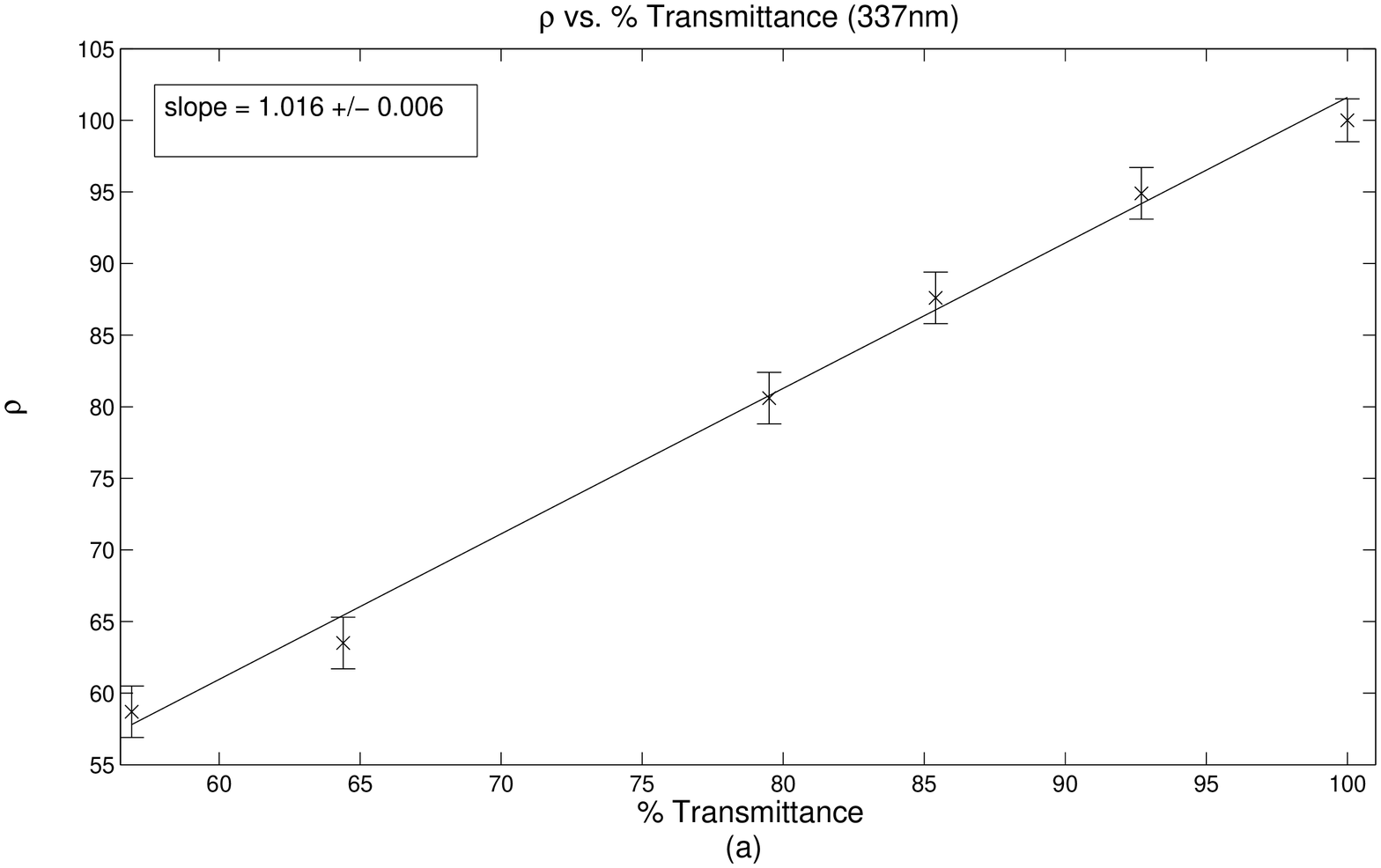,height=6.5cm,width=0.8\textwidth}
\epsfig{file=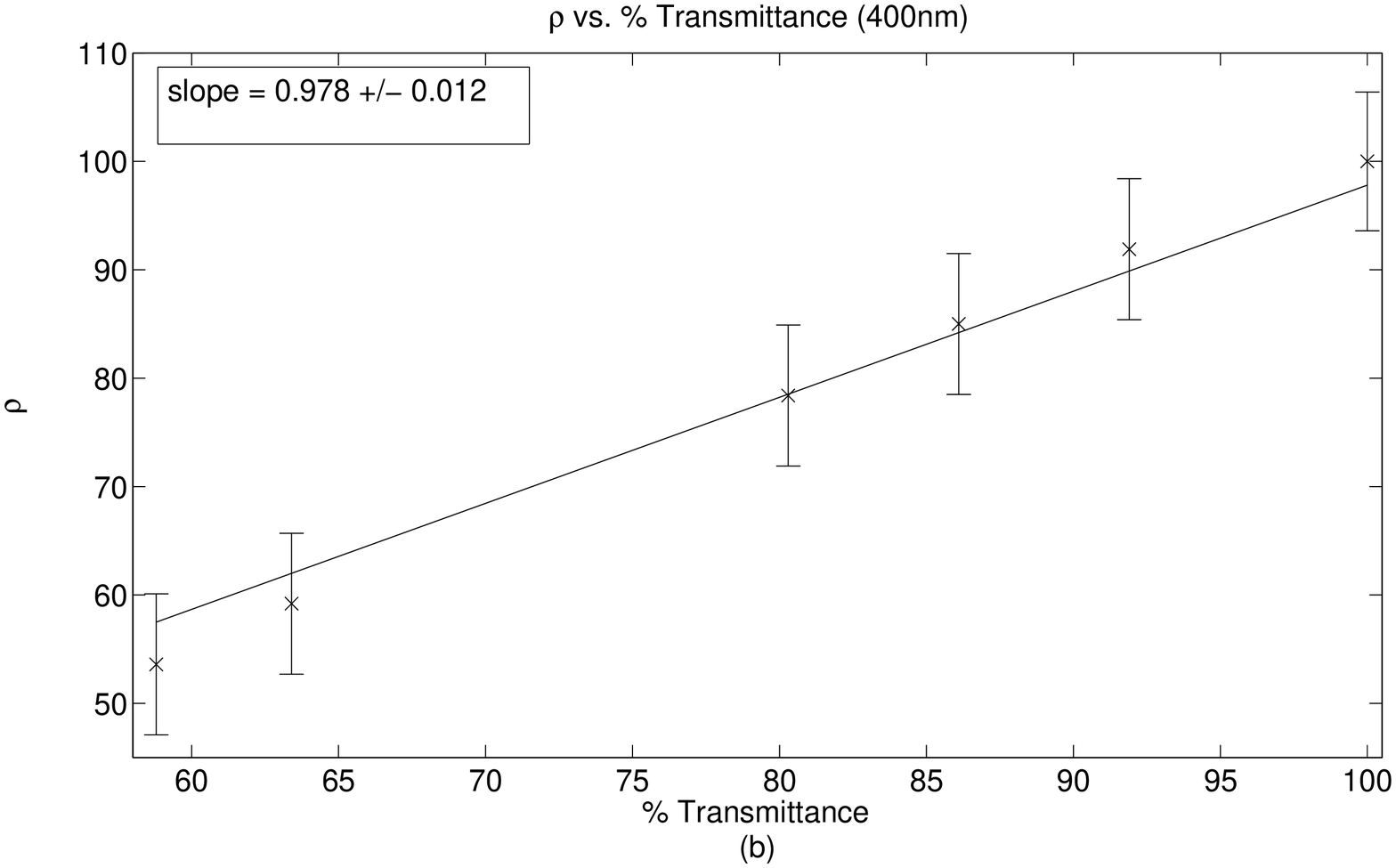,height=6.5cm,width=0.8\textwidth}
\epsfig{file=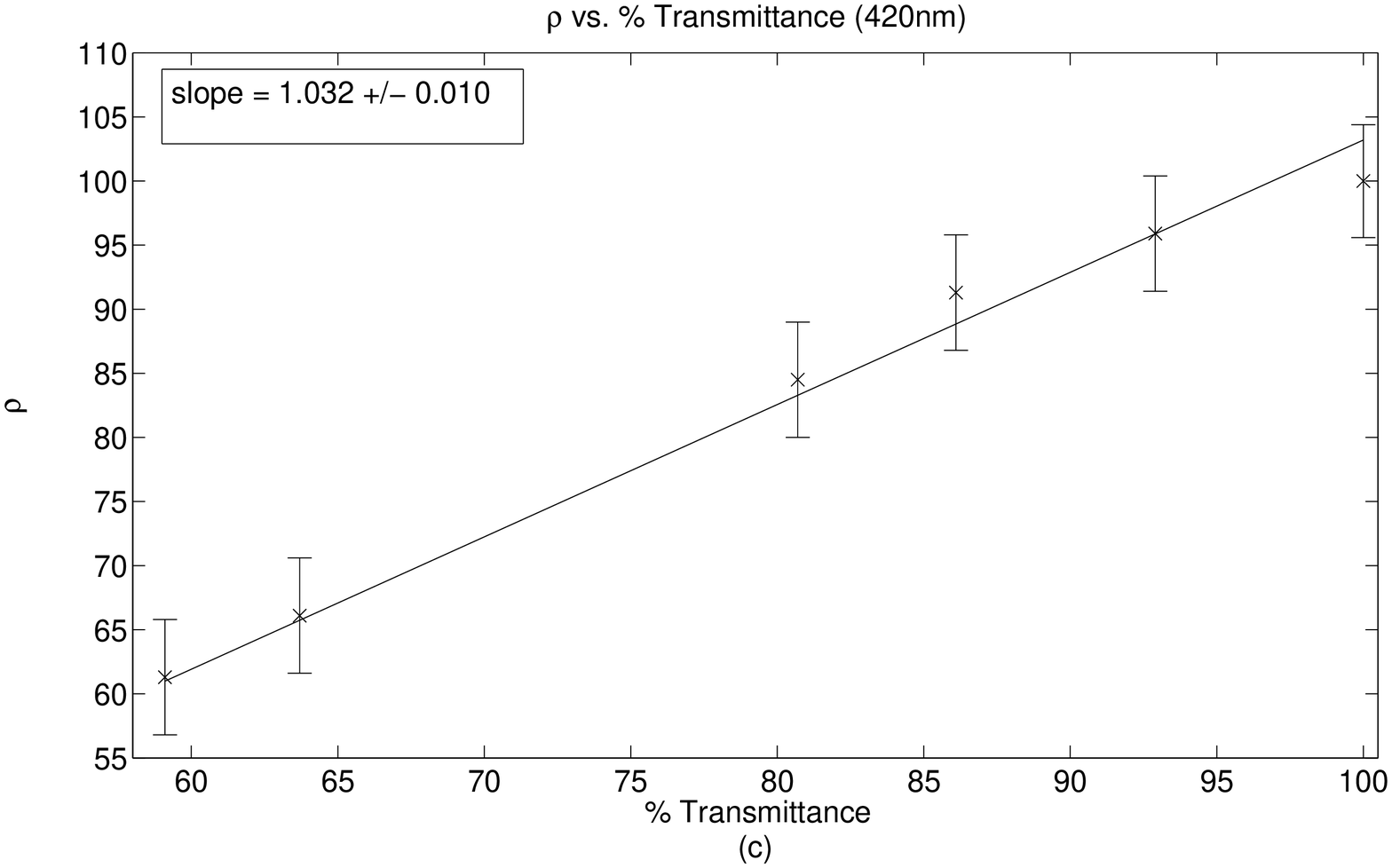,height=6.5cm,width=0.8\textwidth}
\end{center}
\caption{Measurements of light transmission linearity calculated by determining the change in $\rho$ caused by placing neutral density filters of known light transmission in the laser light path for 337nm (a), 400nm (b) and 420nm (c).  The error bars for each graph indicate estimated sources of error excluding the transmission linearity.  The linearity error is taken to be the difference in the slope of the fitted line from 1.}
\label{f:transmission}
\end{figure}

Lastly, there was an uncertainty associated with the long$-$term stability of our detector.  This uncertainty was quantified by determining the value of $\rho$ after the GdCl$_3$ had been removed from the LTA and pure water re-added.  We were able recover the original pure water baseline to within $\pm$1.0$\%$, $\pm$2.5$\%$ and $\pm$2.0$\%$ for 337nm, 400nm and 420nm wavelengths, respectively.  These values are taken as conservative estimates of the long-term drift. Table 1 lists the estimated uncertainties for our measurements at each wavelength.  
\vspace{5mm}

\begin{table*}[h]
\begin{center}
\begin{tabular}{|c|c|c|r|}
\hline
 & \multicolumn {3} {c|} {Wavelength}\\
\bf {Uncertainty} & \bf {337nm} & \bf {400nm} & \bf {420nm} \\
\hline
\hline
Short$-$Term Stability & 0.6 & 4.4 & 2.0 \\
\hline
Alignment & 1.0 & 2.0 & 2.0 \\
\hline
Long$-$Term Stability & 1.0 & 2.5 & 2.0 \\
\hline
Linearity & 0.8 & 1.1 & 1.6 \\
\hline
\hline
\bf {Total} & \bf {1.7 $\%$} & \bf {5.6 $\%$} & \bf {3.8 $\%$} \\
\hline
\hline
\end{tabular}
\caption{\label{tab:5/tc}Estimated uncertainties associated with the measurement of $\rho$.}
\end{center}
\end{table*}

\section{Results}
\subsection{The Addition of GdCl$_{3}$ to Pure Water}

To ensure instrument stability and determine the long$-$term uncertainty as described in Section 3, a pure-water baseline from which to measure the relative change in transparency was obtained  prior to adding the GdCl$_3$.  

To ensure that this procedure did not effect transparency, we $\it {simulated}$ the addition of GdCl$_3$ by mixing pure water from the mixing tank through the $0.22{\mu}$m and $5{\mu}$m filters, and the UV sterilizer.  The pure water was not sent through the de-ionizer since after the GdCl$_3$ is mixed with water this solution does not go through the de-ionizer.  The pure water was then circulated through the LTA for 3 hours, the same length of time used for mixing the GdCl$_3$ with water.  No significant change was observed at any of the three wavelengths immediately after the $\it {simulated}$ chemical addition.  

1.23Kg of GdCl$_3$$\cdot$6H$_2$O was added and mixed with 610L of pure water contained in the mixing tank and stainless steel piping to give a concentration of $0.2\%$ GdCl$_3$ by weight.  The procedure for adding the GdCl$_3$ to the LTA was as follows:
\begin{enumerate}
\item The GdCl$_3$ was initially mixed in a 1 liter beaker of pure water.
\item The GdCl$_3$ - water solution was then added to the poly-propylene mixing tank and mixed using the stainless steel motor-driven stirrer for 10$-$15 minutes.
\item The system valve-alignment was changed to by-pass the de-ionizer.
\item The GdCl$_3$ - water solution was circulated from the mixing tank through the $0.22{\mu}$m and $5{\mu}$m filters and UV sterilizer.
\item The inlet and outlet valves to the LTA were opened and the GdCl$_3$ - water solution was circulated through the LTA for approximately three hours (roughly 3 turn-over times).
\item The LTA inlet and outlet valves were closed and the water pump was secured.
\end{enumerate}

The results of measurements obtained after addition of GdCl$_3$ observed for roughly a 2 day period are shown Figure 7.  Two results are clear.  First, the decrease in $\rho$ over time was consistent with a linear decrease.  In fact, since the GdCl$_3$-water solution remains undisturbed in the LTA after it is mixed, this result suggests a proportionality between decreasing transparency and a change in water quality due to increasing water exposure (with time) to the stainless steel LTA surface.  This point should be stressed: following the mixing of the GdCl$_3$ with the water, the LTA is isolated.  The GdCl$_3$ solution has no contact with anything except the LTA surface, the PVC baffles in the LTA and the acrylic LTA windows.

\begin{figure}
\begin{center}
\epsfig{file=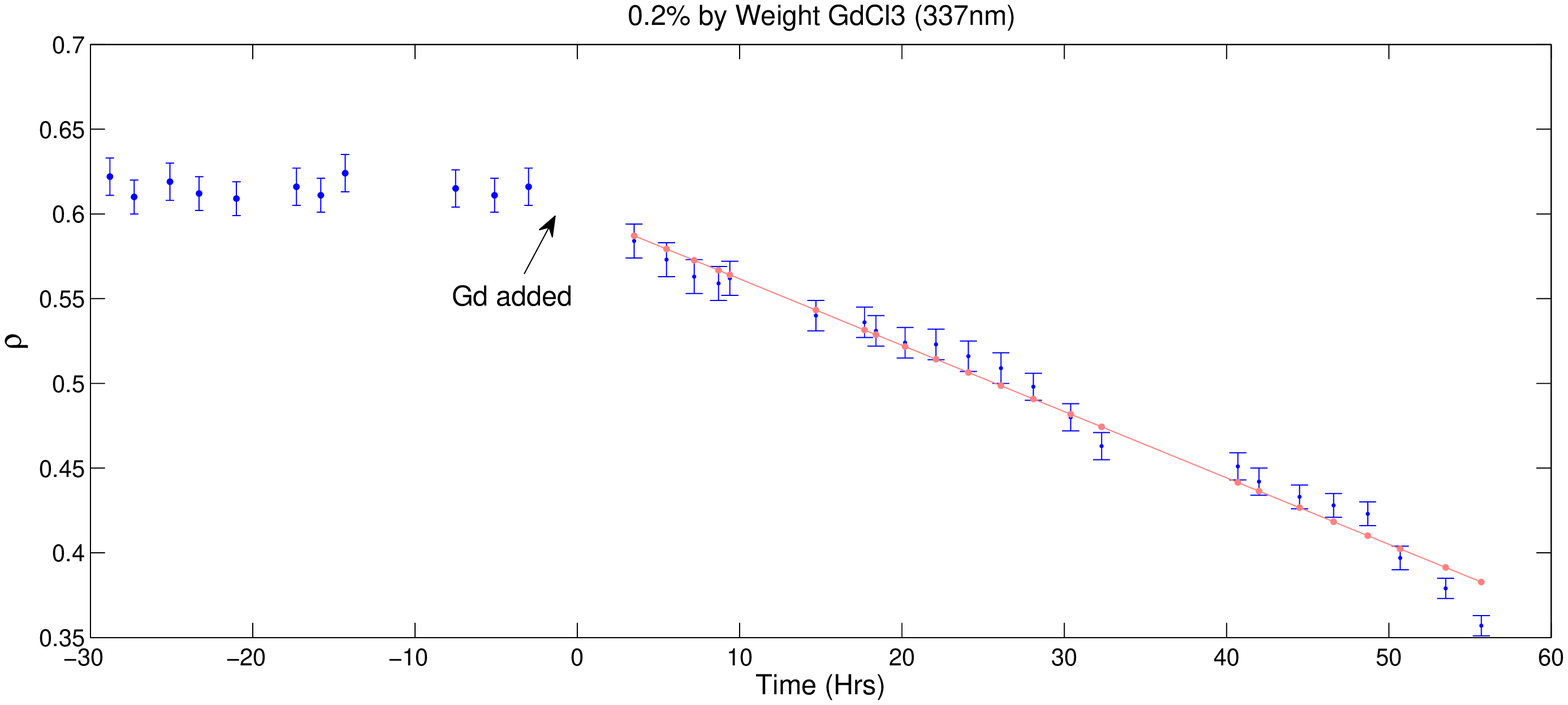,height=6.5cm,width=0.8\textwidth}
\epsfig{file=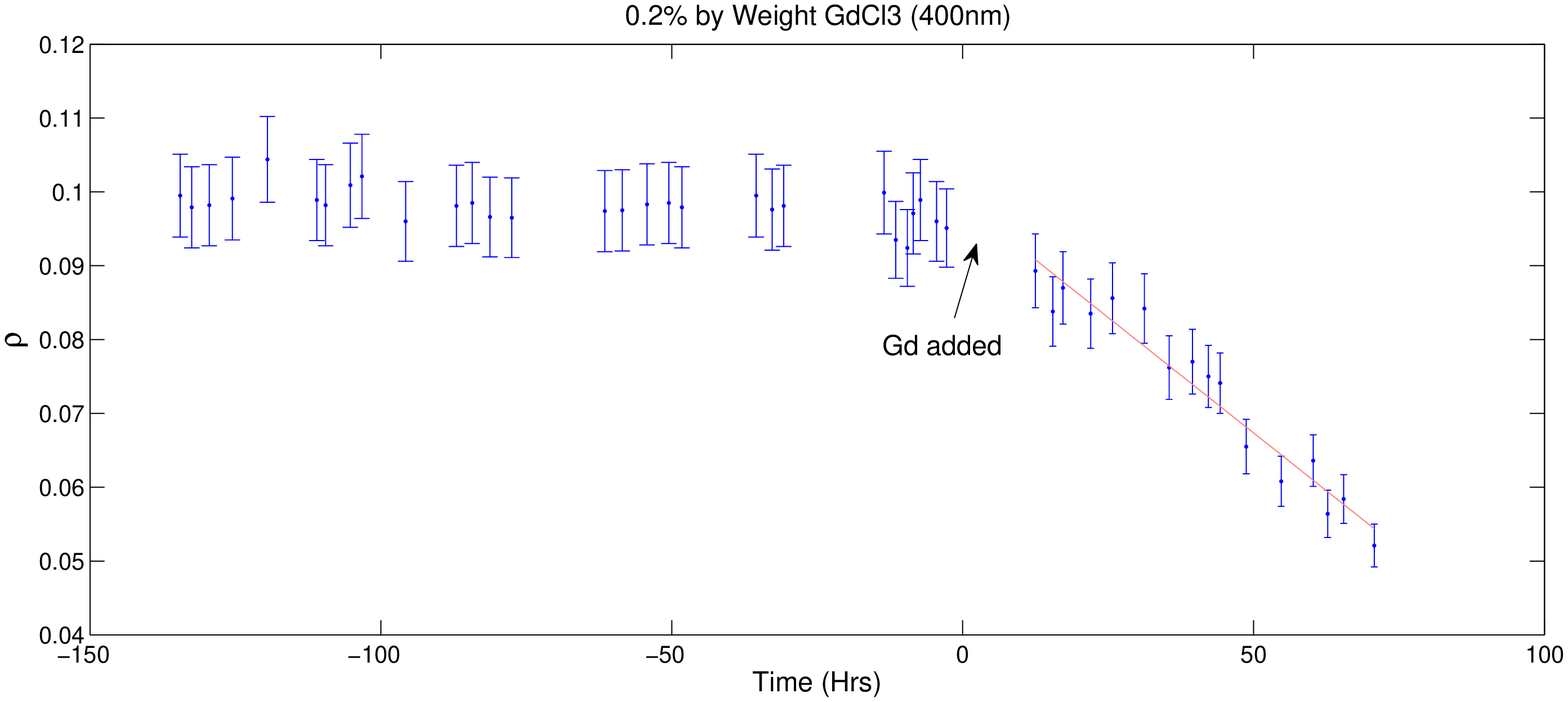,height=6.5cm,width=0.8\textwidth}
\epsfig{file=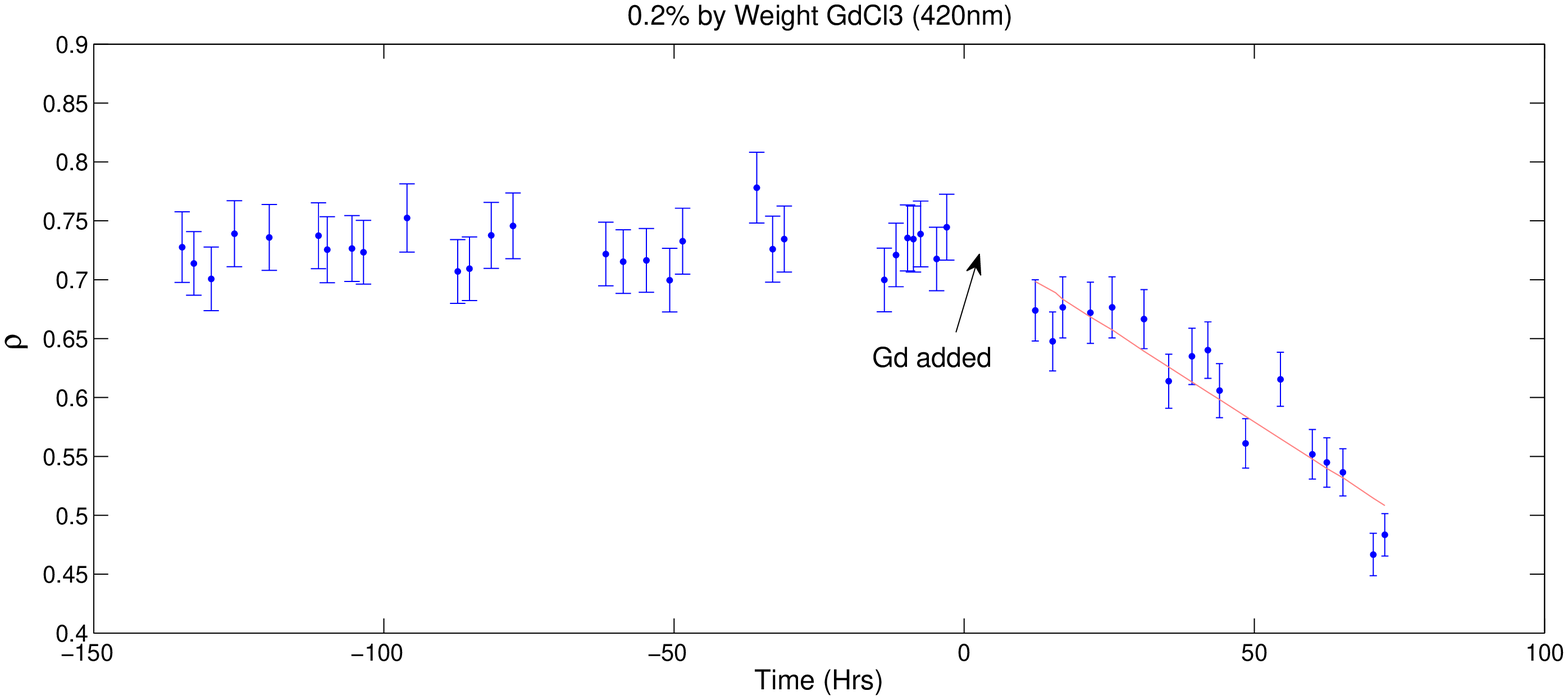,height=6.5cm,width=0.8\textwidth}
\end{center}
\caption{Decrease in transparency versus time resulting from addition of 0.2$\%$ GdCl$_3$ in pure water for 337nm (a), 400nm (b) and 420nm (c).  The red line shows the least squares best fit to the data after addition of the GdCl$_3$.}
\label{f:gdadded}
\end{figure}

Table 2 provides the fitted slope and the y-intercept $\rho$ values (for T $=$ 0) for all three wavelengths.  The last column of Table 2 indicates the level of dissolved oxygen at the time that the GdCl$_3$ was mixed with the pure water.  The slope and intercepts in Table 2 correspond to the following values for $\Delta$$\alpha$ at the 90$\%$ confidence level: $\Delta$$\alpha_{337}$ = 3.4 $\pm$ 28.0 $\times$$10^{-4}$ m$^{-1}$, $\Delta$$\alpha_{400}$ = 5.3 $\pm$ 66.0 $\times$$10^{-4}$ m$^{-1}$ and $\Delta$$\alpha_{420}$ = 1.2 $\pm$ 53.0 $\times$$10^{-4}$ m$^{-1}$.
\vspace{5mm}

\begin{table*}[!h]
\begin{center}
\begin{tabular}{|c|c||c|c|c||c|}
\hline \hline
$\lambda$  & pure water & slope & intercept & reduced & O$_2$ \\
nm & mean  &  $\times$10$^{-4}$ (hr$^{-1}$) & & $\chi^{2}$ & (ppm)\\
\hline
 337	  & 0.62 & -42.0 $\pm$1.7 & 0.619 $\pm$0.006 & 0.60 & 0.90 \\
 400	  & 0.10 & -6.2 $\pm$0.4 & 0.100 $\pm$0.002 & 0.33 & 0.15 \\
 420	  & 0.73 & -30.0 $\pm$2.9 & 0.744 $\pm$0.019 & 0.61 & 0.15 \\
\hline
\end{tabular}
\caption{The Fit parameters (slope and intercept) for the linear decrease in $\rho$ observed after the addition of GdCl$_3$ for the three measured wavelengths. }
\end{center}
\vspace{5mm}
\end{table*}
At all three wavelengths, the fitted line intercepts the t = 0 axis at a value consistent with the pure water baseline $\rho$ value.  This provides strong evidence that the addition of the GdCl$_3$ alone does not instantaneously decrease water transparency.  Rather, it suggests that the drop in $\rho$ results from the introduction of impurities to the GdCl$_3$ water solution from it's exposure to the walls of the stainless steel pipe.

\subsection{Change in Transparency due to the Presence of Iron in Water}
As discussed above, measurement of $\rho$ made after the addition of GdCl$_3$ indicates that the
addition alone does not cause a direct loss of transparency and that the loss of transparency over time is linear.  These results suggest that exposure of the solution to the surface of the stainless steel LTA may be a source of the decrease in $\rho$.  Clearly, one source of potential contamination is Iron (Fe) since it is a strong absorber of UV.  

To investigate the concentrations of iron required to reduce transparency in our apparatus, small amounts FeCl$_3$ were added to pure water in the mixing tank after by-passing the DI and removing the filters in the water purification system.  The filters were removed from the system to ensure that any iron in the water would not accumulate on the filters.  Table 3 shows the change in $\rho$ due to the addition of 14ppb and 28ppb of FeCl$_3$ in pure water at a wavelength of 337nm. 
\vspace{5mm}  

\begin{table*}[!h]
\begin{center}
\begin{tabular}{|c|c|c|}
\hline \hline
pure water value  & 14ppb FeCl$_3$ in water & 28ppb FeCl$_3$ in water \\
\hline
0.901 $\pm$ 0.018	  & 0.355 $\pm$ 0.018   	& 0.156 $\pm$ 0.008  \\
\hline
\end{tabular}
\caption{The change in $\rho$ resulting from the addition of FeCl$_3$ to pure water}
\end{center}
\end{table*}
\vspace{5mm}
As seen in the table, the change in $\rho$ due to the addition of only 14ppb of FeCl$_3$ to pure water results in a reduction in $\rho$ to about of 40$\%$ of the pure water value while the addition of 28ppb of FeCl$_3$ to water drops $\rho$ to 16$\% $ of the pure water value.

It is clear that the presence of extremely small amounts of FeCl$_3$ significantly reduce water transparency.  

\section{Conclusion and Future Work}

We have shown that GdCl$_3$ is problematic for use as a dopant for detectors lined with
stainless steel (e.g. Super-Kamiokande) due to it's effects on water transparency.  At  concentrations of 0.2$\%$ by weight, the transparency of the GdCl$_3$ doped water decreases rapidly over time scales of a few days for all three UV wavelengths tested.  However, since the addition of GdCl$_3$ by itself does not reduce transparency, it may be suitable for detectors made of non-corrosive materials. 

The significant benefits of using gadolinium doped WCDs suggest that additional investigation is required to test the effect of GdCl$_3$ on other materials.  In this vein, we plan additional tests to examine the effect on transparency of materials such as polyethylene jacketed steel and acrylic.  Additionally, other chemicals such as Gd$_2$(SO$_4$)$_3$ and Gd(NO$_3$)$_3$ should be tested for suitability as a gadolinium WCD dopant.

\section{Acknowledgment}  

The authors gratefully acknowledge useful discussions with Henry Sobel, William Kropp, Mark Vagins and Michael Smy.  This work was performed under the auspices of the U.S. Department of Energy by Lawrence Livermore
National Laboratory under Contract {DE-AC52-07NA27344} and was supported by Livermore Directed Research Program funding (LDRD 06-FS-011) and U.S. Department of
Energy grants {DE-GF02-91ER40674} and {DE-FG02-91ER40617}.

\end{document}